\documentclass[aps,pra,twocolumn,floats,showpacs,superscriptaddress]{revtex4-2}
\usepackage{graphicx,amsmath,amsfonts,amssymb,upgreek,txfonts,color}
\usepackage[colorlinks,linkcolor=blue,citecolor=blue,urlcolor=blue,breaklinks=true]{hyperref}
\usepackage{color, colortbl}
\definecolor{lightgreen}{rgb}{0.88,1,1}
\begin{document}

	\title{Enhancing Weak magnetic field sensing of cavity-magnon system with dual frequency modulation}

	\author{Zheng Liu, Yu-qiang Liu, Zi-yi Mai, Yi-jia Yang, Nan-nan Zhou and Chang-shui Yu}
	\email{Electronic address: ycs@dlut.edu.cn}
	\address{School of Physics, Dalian University of Technology, Dalian 116024,
		P.R. China}
	\date{\today}

	\begin{abstract}
 The crucial limitation of improving the sensitivity of the detection of weak magnetic fields is the unavoidable measurement noise.  In this paper, we propose a scheme to achieve precise sensing robust against additional noise by employing a dual-frequency bias field modulation within a cavity magnon system.  We find that the anti-rotating wave term can amplify the signal of the detected magnetic field, but this amplification effect must coexist with the rotating wave term.  In particular, by the bias field modulation, we find the robustness against cavity field thermal noise is substantially enhanced, quantum noise and cavity field thermal noise is greatly reduced, and the external magnetic field signal is amplified, thereby improving the weak magnetic field sensing system's sensitivity. Compared with the previous scheme, our scheme requires neither an ultra- or deep-strong coupling mechanism nor the suppression of the additional noise by increasing the electromagnetic cooperativity. Our scheme could provide a valuable candidate for weak magnetic field sensing.
	\end{abstract}

	\maketitle

	\section{\label{sec1}Introduction}

    The precise measurement of weak magnetic fields is currently a major topic of interest in both theory and experiment \cite{PhysRevApplied.19.044044,zhang,PhysRevB.99.214415,PhysRevLett.108.120801,https://doi.org/10.1002/adma.201401144,Li:18,PhysRevA.103.062605}  which is widely applied in geophysics \cite{Stolz_2021}, biology \cite{s20061569,10.1116/5.0025186}, and dark matter search \cite{PhysRevD.99.075031}. Many magnetometers with certain operating frequencies and environments, such as superconducting quantum interference magnetometers (SQUIDs) \cite{10.1063/5.0065790}, atomic magnetometers \cite{PhysRevLett.130.023201}, and nitrogen-vacancy centers magnetometers \cite{PhysRevApplied.11.034029}, have been extensively studied. How to realize a weak magnetic field sensing with high sensitivity, wide frequency range, and noise immunity is attracting increasing interest.
	
	Additional noise is the core factor limiting the detection sensitivity in quantum sensing \cite{RevModPhys.82.1155}, which, distinguished from probe noise, includes the quantum noise and thermal noise of the whole system except the probe. Over the past decade,  many approaches are proposed to reduce the quantum noise without thermal noise explicitly taken into account. For example, quantum noise is reduced in precision measurement and break the standard quantum limit (SQL) in optomechanical force sensing such as coherent quantum noise cancellation (CQNC) \cite{PhysRevA.92.043817,10.3389/fphy.2023.1142452,Motazedifard_2016,moller2017quantum,vitali,PhysRevA.89.053836}, non-Markovian regime \cite{Zhang_2017}, and squeezed cavity field \cite{PhysRevLett.115.243603,zhao2020weak,aasi2013enhanced,galaxies10020046,PhysRevLett.129.121103}, and so on.  The mechanical oscillator's sensitivity to external forces and the unique optical readout in the cavity optomechanical system enables it to naturally detect weak force signals \cite{RevModPhys.86.1391,PhysRevA.90.043848}. Therefore, if some systems have highly sensitive probes to measure external magnetic fields and the corresponding readout device, these systems can also be used for magnetic field sensing. and the above scheme for reducing quantum noise is also applicable to weak magnetic sensing.
  
	Cavity magnon systems, composed of microwave cavity and yttrium iron garnet (YIG) sphere, have recently gained significant attention in addition to the cavity optomechanical systems \cite{PhysRevLett.113.156401,ZARERAMESHTI20221,doi:10.1126/sciadv.1501286,Lachance-Quirion_2019,PhysRevLett.124.053602,PhysRevLett.104.077202,PhysRevD.99.075031}. The cavity-magnon system can serve as a feasible platform to investigate quantum magnetic field sensing in the field of quantum precision measurement. The reason is, on one hand, the YIG sphere is an excellent ferromagnetic material that produces a low excitation Magnon mode (Kittel mode) \cite{PhysRev.110.836} with high spin density, low decay rate \cite{PhysRevApplied.2.054002}, and high-frequency tuning \cite{PhysRevB.96.094412}, which is beneficial to highly sensitive magnetic field sensing, and the microwave cavity, on the other hand, facilitates microwave readout and can achieve strong coupling or even ultra-strong coupling with the YIG sphere \cite{PhysRevLett.111.127003,PhysRevLett.113.083603,PhysRevB.93.144420,10.1063/1.4941730}, making them ideal for various quantum information processing applications, such as the preparation of macroscopic Schrödinger cat states \cite{PhysRevLett.127.087203}, steady-state magnon entanglement \cite{PhysRevLett.121.203601,Zhang:22,Yang:20,PhysRevA.105.063704,10.1063/5.0015195}, and blockade of magnon \cite{PhysRevB.100.134421,PhysRevA.101.063838}. However, the sensing of weak magnetic fields using the cavity-magnon system remains challenged due to the presence of additional noise, i.e., quantum noise and microwave cavity field thermal noise. Even though the ultra-strong or deep-strong coupling allows the interaction of the anti-rotating wave, which can effectively suppress quantum noise and cavity field thermal noise below the SQL and amplify the signal \cite{PhysRevA.103.062605}, realizing the ultra-strong or deep-strong coupling \emph{per se} is also more challenging than the realization of the strong coupling, and additionally, what role the rotating wave and anti-rotating wave interactions play in weak magnetic field sensing remains unclear. 
	
	In this paper, to sense weak magnetic field signal we introduce a dual-frequency bias magnetic field modulation to the cavity magnon system with the YIG sphere coupling with a microwave cavity through dipole-dipole interaction \cite{PhysRevB.107.L140405}. The external magnetic field interacting with the YIG sphere affects the microwave output spectrum which can be measured by the cavity field phase quadrature detection method. It is shown that the dual-frequency bias magnetic field modulation can realize the anti-rotating wave interaction through strong coupling regime instead of ultra-strong or deep-strong coupling regime. It can especially control the proportion of rotating and anti-rotating wave terms in the system which can reveal the roles of each type of interaction, by which we find that the anti-rotating wave term can amplify the detected field signal, but this amplification effect must coexist with the rotating wave term. We show that in our scheme, there exists a noise-resistant frequency band tolerating additional noise if selecting appropriate parameters.  In particular, the sensitivity can be improved and the system's response to weak magnetic fields can be amplified in this frequency band. Compared with the scheme achieving uncontrolled anti-rotating wave terms under ultra-strong or deep-strong coupling mechanisms \cite{PhysRevA.103.062605}, we find that our scheme can achieve additional noise suppression of the same order of magnitude under ultra strong or even deep strong coupling mechanisms. This only requires the cavity magnetic strong coupling mechanism, which is of course more conducive to experimental implementation. Moreover, we can foresee that when our scheme is in the super strong coupling mechanism or deep strong coupling mechanism, the additional noise will be suppressed to a greater extent. The rest of the paper is organized as follows.  In Section \ref{sec2}, we introduced the specific model of our dual frequency modulation magnetic field sensing scheme and the derivation of Hamiltonian. In Section \ref{sec3}, we analyses the dissipative dynamics of the system and provided an analytical expression for the phase orthogonal output spectral density used to measure sensing performance. In Section \ref{sec4}, we analyze the roles played by the anti-rotating term and anti-rotating wave term in weak magnetic sensing, and evaluated the response performance of our weak magnetic sensing scheme to external detection magnetic fields and the suppression of additional noise. In addition, the superiority of this scheme was demonstrated through comparison with previous schemes. The discussion and conclusions are given in Section \ref{sec5}.

	\section{\label{sec2} sensing system and Hamiltonian}
	
\begin{figure}
		\includegraphics[width=1.0\columnwidth]{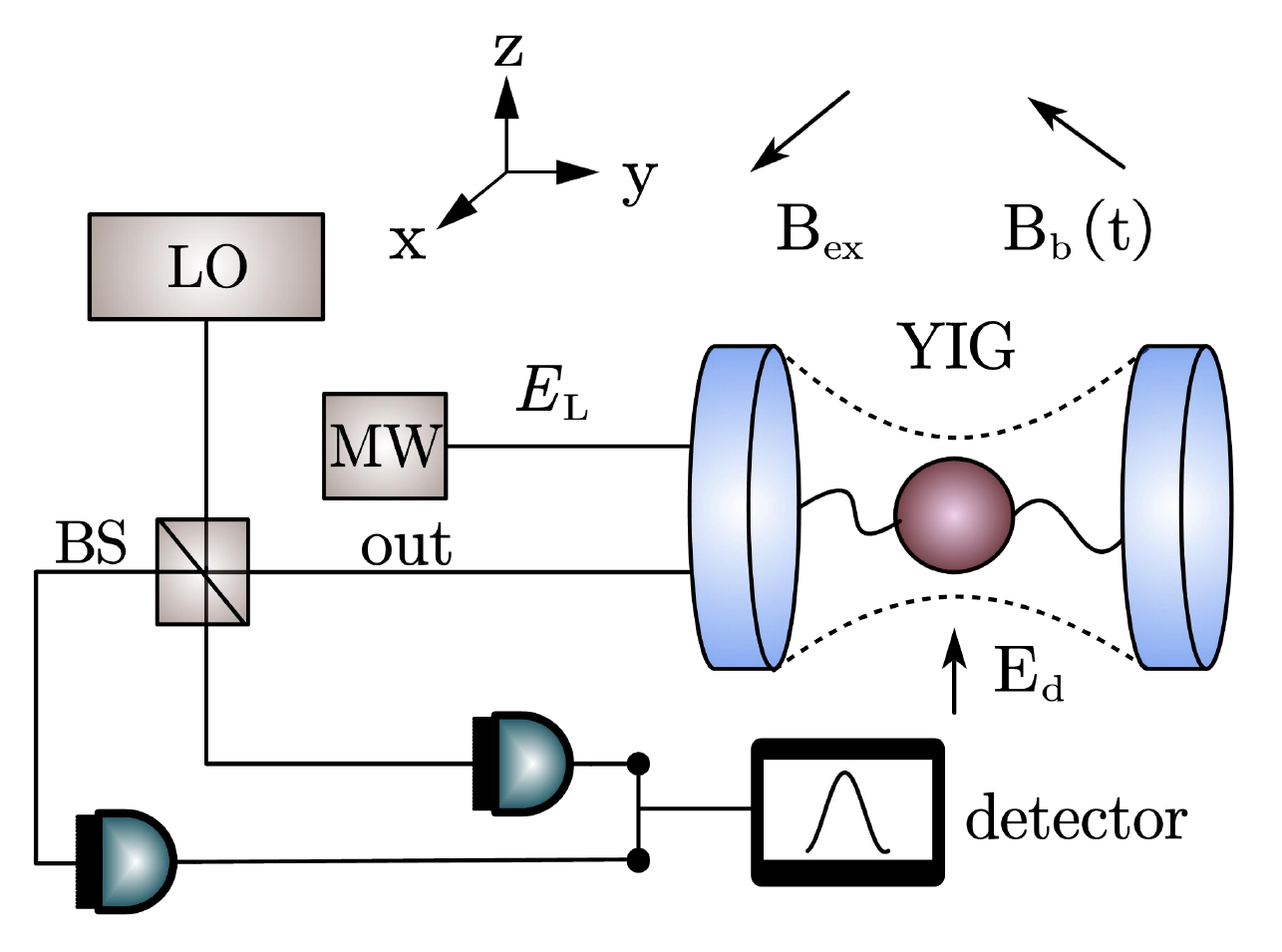}
		\caption{The schematic diagram of weak magnetic field sensing system. The YIG sphere is placed at the maximum value of the field generated by the microwave cavity, the time-dependent dual frequency modulation bias magnetic field $B_b(t)$ is along the x-z direction to regulate the frequency of the magnon mode, and the cavity field and magnon mode are driven by two classical pump fields, whose driving amplitudes are $E_L$ and $E_d$ respectively. Detected magnetic field $B_ {ex}$ assumes along the x direction. The information of the detected magnetic field is determined by the information of the output cavity field, and its output is analyzed by a balanced homodyne detector. }
		\label{Fig1}
	\end{figure}

	 Our system consists of a YIG sphere and a microwave cavity with the YIG sphere acting as a probe for magnetic field sensing and the output of the microwave cavity acting as a readout for external magnetic field information. The diagrammatic sketch is shown in Fig. \ref{Fig1}. In this model,  the coupling between the microwave cavity mode and the magnon mode is generated by the magnetic dipole-dipole interaction and the frequency of the magnon mode can be tuned by the external dual-frequency modulated field $B_b(t)=B_{b0}+\sum_{i=1,2}B_{bi}\cos(\omega _it+\phi_i)$, and the two modes are respectively pumped by two different semi-classical coherent pump fields.  Thus the Hamiltonian of the system reads	
	\begin{eqnarray}\label{1}
		\hat{H}&=&\hbar\omega _a\hat{a}^{\dagger}\hat{a}+\hbar[\omega _m+\sum_{i=1}^2{\lambda _i\omega _i\cos ( \omega _it+\phi_i )}]\hat{m}^{\dagger}\hat{m}\nonumber\\&&+\hbar g( \hat{a}+\hat{a}^{\dagger} ) ( \hat{m}+\hat{m}^{\dagger} )\nonumber+i\hbar E_d( \hat{m}^{\dagger} e^{-i\omega_d t}-\hat{m}e^{i\omega_d t})\nonumber\\
		&&+i\hbar E_L( \hat{a}^{\dagger} e^{-i\omega_L t}-\hat{a}e^{i\omega_L t})-\hbar \epsilon B_{ex}(t) (\hat{m}+\hat{m}^{\dagger}).
	\end{eqnarray}
		The first two terms of the Hamiltonian represent the free Hamiltonian of the cavity mode and the magnon mode, where $\hat a,\hat m$ $(\hat a^{\dagger },\hat m^{\dagger })$ are the bosonic annihilation (creation) operators of cavity mode and magnon mode, $\omega_a$, $\omega_m$ are the resonant frequencies of the cavity field and the magnon mode, $\omega_i$, $\phi_i$ $(i=1, 2)$ denote the two frequencies and phases of the dual-frequency modulated bias magnetic field, respectively, and $\lambda_i$ $(i=1, 2)$ is the coupling coefficient between the dual-frequency modulated bias magnetic field and the magnon mode. Due to the bias field, the frequency of the magnon mode of the YIG sphere can be modulated, i.e., $\omega_m=\gamma B_{b0}$, $\lambda_i=\gamma B_{bi}/\omega_i$. The third to the fifth terms are the dipole-dipole interaction between the cavity field mode and the magnon mode, the classical driving of the cavity field, and the driving of the magnon mode, where $g=\frac{\gamma B_0}{2}\sqrt{5N}$ is the coupling coefficient between the cavity mode and the magnon mode, $B_0$ is the amplitude of the microwave field. $E_L =\sqrt{\frac{2P_{L}\kappa_a}{\hbar\omega_L}} $, $\omega_L$ ($E_d=\frac{\gamma B_d}{4}\sqrt{5N}$, $\omega_d$) \cite{PhysRevLett.121.203601} is the amplitude and frequency of the microwave pump field (magnon mode driving field), $P_{L}$ represents the input power of the microwave cavity field, $B_d$ is the magnetic induction intensity of the magnon driving field. According to the experiment, the gyro-magnetic ratio $\gamma/2\pi = {28} ~\mathrm{GHz/ T}$ \cite{doi:10.1126/sciadv.1501286}. The last term represents the coupling of the probed magnetic field to the YIG sphere, which can be obtained from the  Holstein-Primakkoff transformations \cite{PhysRev.58.1098}, where $\epsilon =\frac{\gamma}{2} \sqrt{5N}$ is the coupling coefficient between the detected magnetic field and the YIG sphere. It is worth noting that $\omega_ i$ represents the frequency of the external driving field, and its magnitude is in the range of $10 ^ {1} \rm GHz$, the value of $\lambda_ i$ depends on the ratio of the product of the external modulation field $B_{bi}$ and the rotational magnetic ratio $\gamma$ to the frequency of the modulation field. According to existing experimental conditions, The magnitude of B is $10 ^{ 0} \rm T$, so it is reasonable for $\lambda_i$ to range from $10 ^ {-1}$ to $10 ^ {0}$.

	Considering a special unitary transformation operation $\hat{U}( t )=\exp {[ -i( \omega _L\hat{a}^{\dagger}\hat{a}+\omega _d\hat{m}^{\dagger}\hat{m} )] t }\cdot\exp{[-i\sum_{i=1}^2{\lambda _i\sin ( \omega _it +\phi_i)\hat{m}^{\dagger}\hat{m}}]} $ and the Jacobi-Anger expansions
	$e^{i\lambda\sin x}=\sum_{m=-\infty}^{\infty}{J_m(\lambda)}e^{imx}$
\cite{https://doi.org/10.1002/andp.202100421,Chao:21}, the Hamiltonian becomes 
	\begin{eqnarray}\label{2}
		\hat{H}^{\prime}&=&\hbar\Delta _a\hat{a}^{\dagger}\hat{a}+\hbar\Delta _m\hat{m}^{\dagger}\hat{m}-\hbar \epsilon B_{ex}(t) (\hat{m}e^{-i\omega_d t}+\hat{m}^{\dagger}e^{i\omega_d t})\nonumber\\&&+\hbar( g_{1}\hat{m}+g_{2}\hat{m}^{\dagger})\hat{a}+\hbar(g_{1}\hat{m}^{\dagger}+g_{2}\hat{m})\hat{a}^{\dagger}\nonumber\\&&+i\hbar E_L(\hat{a}^{\dagger}-\hat{a})+i\hbar E_d(\hat{m}^{\dagger}-\hat{m}),
	\end{eqnarray}
	where $\Delta_a = \omega_a-\omega_L$ and $\Delta_m=\omega_m-\omega_d$ are the detunings of the cavity mode and the magnon mode relative to their respective driving field frequencies,
	the time-dependent coupling strength $g_1$ and $g_2$  can be explicitly given as
\begin{eqnarray}\label{3}
	\frac{g_{1}}{g}&=\sum_{m_{1,2}=-\infty}^{\infty}{J_{m_1}(\lambda_1)}J_{m_2}(\lambda_2) e^{-i(\omega _L+\omega _d+m_1\omega_1+m_2\omega_2)t-i( m_1\phi _1+m_2\phi _2 )},\nonumber\\
	\frac{g_{2}}{g}&=\sum_{n_{1,2}=-\infty}^{\infty}{J_{n_1}(\lambda_1)}J_{n_2}(\lambda_2) e^{-i(\omega _L-\omega _d-n_1\omega _1-n_2\omega _2)t+i( n_1\phi _1+n_2\phi _2 )}.
\end{eqnarray}
Let's focus on the strong coupling regime \cite{PhysRevLett.113.156401}, i.e. $\kappa_a$, $\kappa_m<g \ll \omega_a$, $\omega_m$, and let $\omega_1=\omega _d-\omega _L$, $\omega_2=\omega _L+\omega _d$. If $| ( 1-m_1+m_2 ) \omega _a+( 1+m_1+m_2 ) \omega _d |\ll g$,
$| ( 1+n_1-n_2 ) \omega _a-( 1+n_1+n_2 ) \omega _d |\ll g$, the high-frequency oscillation term can be safely neglected, so we can only consider $m_ 1=0, m_ 2=-1$, $n_ 1=-1$ and $ n_ 2=0 $.  Thus one can arrive at the final Hamiltonian as
\begin{eqnarray}\label{4}
	\hat{H}_{f}&=&\hbar\Delta _a\hat{a}^{\dagger}\hat{a}+\hbar\Delta _m\hat{m}^{\dagger}\hat{m}
	+\hbar( g_1\hat{m}e^{i\phi_2}+g_2\hat{m}^{\dagger}e^{-i\phi_1} )\hat{a}\nonumber\\&&
	+\hbar( g_1\hat{m}^{\dagger}e^{-i\phi_2}+g_2\hat{m}e^{i\phi_1} )\hat{a}^{\dagger}+i\hbar E_L(\hat{a}^{\dagger}-\hat{a})\nonumber\\&&+i\hbar E_d(\hat{m}^{\dagger}-\hat{m})-\hbar \epsilon B_{ex}(t) (\hat{m}e^{-i\omega_d t}+\hat{m}^{\dagger}e^{i\omega_d t}),
\end{eqnarray}
where the cavity magnetic coupling strength can be simplified as
\begin{align}\label{5}
		 g_{1}= g J_0( \lambda _1 ) J_{-1}( \lambda _2 ),
	     g_{2}= g J_0( \lambda_2 )J_{-1}( \lambda _1 ) .
\end{align}
Eq. (\ref{5}) indicates that  $g_1$ and $g_2$ can be adjusted by $\lambda_1$ and $\lambda_2$. Since $g_1$ reflects the dual mode squeezing interaction corresponding to the anti-rotating wave term, and $g_2$ reflects the beam splitter type interaction corresponding to the rotating wave term, one can freely adjust the ratio of the rotating wave term and the anti-rotating wave term, which is of great advantage to quantum information processing. In addition, we do not need ultra-strong coupling or deep strong coupling mechanisms for the anti-rotational wave term, which is also easier to achieve in experiments \cite{10.1063/1.4941730,hu2023twotone}.

It is worth emphasizing that the selection of frequencies is essential for achieving the desired interactions and behaviors in this system.  The conditions ($\omega_{1}$ = $\omega_{d}$ - $\omega_{L}$) and ($\omega_{2}$ = $\omega_{L}$ + $\omega_{d}$) are important for eliminating time-dependency in the coupling strengths ($g_1$) and ($g_2$). These conditions facilitate the use of a special unitary transformation employing the Jacobi-Anger expansion, transforming a time-dependent problem into a time-independent one. By setting ($\omega_{1}$ = $\omega_{d}$-$\omega_{L}$) and ($\omega_{2}$ = $\omega_{L}$ + $\omega_{d}$), time-dependent terms in the Hamiltonian are negated, resulting in a time-independent effective Hamiltonian. This simplification is vital for the analytical tractability and comprehension of the system's quantum dynamics. Deviation from these conditions could reintroduce time dependency, complicating the analysis and possibly leading to less efficient mode coupling. This could diminish the effectiveness of the sensing.  In practical applications, these conditions dictate the experimental setup, influencing the choice of frequencies for the applied fields and the design of the cavity and magnon systems to fulfill the required conditions.
	\section{\label{sec3} DYNAMICS OF THE SENSING SYSTEM}
	For simplicity, in the following, we set $\phi_i$ = 0 $(i=1,2)$. Considering the fluctuations and dissipations, the quantum Langevin equation \cite{PhysRevLett.46.1} of the weak magnetic field sensing system can be written as
	\begin{eqnarray}\label{6}
		 \dot{\hat{a}}&=&-i\Delta_a \hat{a}-\frac{\kappa_a}{2}\hat{a}-ig_1 \hat m^{\dagger}-ig_2 \hat{m}+E_L+\sqrt{\kappa_a} \hat{a}_{\rm in}(t),\nonumber\\
		 \dot{\hat{m}}&=&-i\Delta_m  \hat{m}-\frac{\kappa_m}{2} \hat m-ig_1 \hat a^\dag-ig_2 \hat{a}+E_d+\sqrt{\kappa_m} \hat{m}_{\rm in}(t)\nonumber\\
		&&+i\epsilon B_{ex}(t) e^{i\omega_d t},
	\end{eqnarray}
 where $\kappa_a$ and $\kappa_m$ are the dissipations of the cavity field and the magnon mode, respectively, and $\hat{a}_{\rm in}(t)$ and $\hat{m}_{\rm in}(t)$ are the input noise operators of the vacuum field cavity mode and the magnetic mode, respectively.
 These operators are responsible for the stochastic behaviors of the system due to the coupling of the system to its environment. These noise operators satisfy the commutation relationship of bosonic operators and have zero mean values, i.e.,
 \begin{align}\label{7}
 	&\left\langle \hat a_{\rm in}(t) \hat a_{\rm in}^{\dagger}\left(t^{\prime}\right)\right\rangle= (\bar n_{a }+1)\delta\left(t-t^{\prime}\right),\notag \notag \\& \left\langle \hat a_{\rm i n}^{\dagger}(t) \hat a_{\rm i n}\left(t^{\prime}\right)\right\rangle=\bar n_{a } \delta\left(t-t^{\prime}\right),\notag \notag \\&\left\langle \hat m_{\rm in}(t) \hat m_{\rm in}^{\dagger}\left(t^{\prime}\right)\right\rangle= (\bar n_{m }+1)\delta\left(t-t^{\prime}\right),\notag \notag \\& \left\langle \hat m_{\rm i n}^{\dagger}(t) \hat m_{\rm i n}\left(t^{\prime}\right)\right\rangle=\bar n_{m} \delta\left(t-t^{\prime}\right),
 \end{align}
where $\bar n_a=[
\exp ( \hbar \omega _ { a } / k _ { B } T ) - 1 ] ^ { - 1 } $ ($\bar n_m=[
\exp ( \hbar \omega _ { m } / k _ { B } T ) - 1 ] ^ { - 1 }$) is the thermal occupancy number of the cavity field mode (magnon mode). Based on the strong classical coherent driving of the system, we can write $\hat a=\bar{a}+\delta {\hat{a}}$, $\hat m=\bar{m}+\delta {\hat{m}}$, and ignore high-order fluctuations. So the Langevin equations for the fluctuation operators read
\begin{eqnarray}\label{8}
	\delta \dot{\hat{a}}&=&-i\Delta_a \delta \hat{a}-\frac{\kappa_a}{2} \delta\hat{a}-ig_1  \delta\hat m^{\dagger}-ig_2 \delta\hat{m}+\sqrt{\kappa_a} \hat{a}_{\rm in}(t),\nonumber\\
	\delta \dot{\hat{m}}&=&-i\Delta_m \delta \hat{m}-\frac{\kappa_m}{2} \delta\hat m-ig_1 \delta\hat a^\dag-ig_2 \delta\hat{a}+\sqrt{\kappa_m} \hat{m}_{\rm in}(t)\nonumber\\
	&&+i\epsilon B_{ex}(t) e^{i\omega_d t}.
\end{eqnarray}
To better calculate the phase quadrature component sensing, we rearrange the above equations (\ref{9})  in the matrix form with the corresponding orthogonal fluctuation operators as
\begin{eqnarray}\label{9}
	\dot{\hat{V}}=C\hat{V}+\hat V_{in},
\end{eqnarray}
where $\hat{V}$ = $[\delta\hat{X}_{a}(t),\delta\hat{P}_{a},\delta\hat{X}_{m}
,\delta \hat{P}_{m}]^{T}$, $\hat V_{in}=[\sqrt{\kappa _{a}}\hat x^{\rm in}_{a},\sqrt{\kappa _{a}}\hat p^{\rm in}_{a},\sqrt{\kappa _{m}}\hat x^{\prime\rm in}_{m},\sqrt{\kappa _{m}}\hat p^{\prime\rm in}_{m}]^{T}$, and
\begin{align}\label{10}
	C=\left(
	\begin{matrix}
		{\frac{-\kappa_a}{2}} & {\Delta _{a}} & {0} & {g_2-g_1} \\
		{-\Delta _{a}} & {-\frac{\kappa_a}{2}} & {-(g_1+g_2) } & {0} \\
		{0} & {g_2-g_1} & {-\frac{\kappa_m}{2}} & {\Delta _{m}}  \\
		{-(g_1+g_2) } & {0} & {-\Delta _{m}} & {-\frac{\kappa_m}{2}} \\
	\end{matrix}
	\right),
\end{align}%
with $\delta\hat{X}_{a}=(\hat{a}^{\dagger }+\hat{a})/\sqrt{2}$, $\delta\hat{P}_{a}=(%
\hat{a}-\hat{a}^{\dagger })/\sqrt{2}i$, $\delta\hat{X}_{m}=(%
\hat{a}+\hat{a}^{\dagger })/\sqrt{2}i$, $\delta\hat{P}_{m}=(%
\hat{a}-\hat{a}^{\dagger })/\sqrt{2}i$, $\hat{x}_{a}^{in}=(\hat{a}%
_{in}^{\dagger }+\hat{a}_{in})/\sqrt{2}$, $\hat{p}_{a}^{in}=(\hat{a}_{in}-%
\hat{a}_{in}^{\dagger })/\sqrt{2}i$. It is worth noting that the definition of the quadrature component of the magnon mode has been modified as
	\begin{eqnarray}\label{11}
		\hat{x}^{\prime \mathrm{in}}_m(t)&=&\hat{x}^{\rm{in}}_m-\sqrt{\frac{2}{\kappa_m}}\epsilon B_{ex}(t) \sin (\omega_d t), \label{15}\nonumber\\
		\hat{p}^{\prime \mathrm{in}}_m(t)&=&\hat{p}^{\rm{in}}_m+\sqrt{\frac{2}{\kappa_m}}\epsilon B_{ex}(t) \cos(\omega_d t), 
	\end{eqnarray}
where $\hat{x}_{m}^{in}=(\hat{m}%
_{in}^{\dagger }+\hat{m}_{in})/\sqrt{2}$ and $\hat{p}_{m}^{in}=(\hat{m}_{in}-%
\hat{m}_{in}^{\dagger })/\sqrt{2}i$ are the quadrature component operators of the magnon mode before correction. 

The stability of the system can be guaranteed by the Routh-Hurwitz stability criterion \cite{PhysRevA.35.5288} which requires the matrix $C$ satisfying
\begin{align}\label{12}
	&{H}_{3}>0, {H}_{3} {H}_{2}-{H}_{1}>0, \nonumber\\ &{H}_{3} {H}_{2} {H}_{1}-({H}_{1}^{2}+{H}_{3}^{2} {H}_{0})>0,
\end{align}
\begin{widetext}
with
\begin{eqnarray}\label{13}
	{H}_{3}&=&\kappa_a +\kappa_ m,  {H}_{2}=2 (g_{2}^{2}- g_{1}^{2})+\Delta_{a}^{2}+\Delta_{m}^{2}+\frac{\kappa_{a}^{2}}{4}+\frac{\kappa_{m}^{2}}{4}+\kappa_{a} \kappa_{m}, \nonumber\\ {H}_{1}&=&-{g}_{1}^{2} \kappa_{\mathrm{a}}+\mathrm{g}_{2}^{2} \kappa_{\mathrm{a}}+\Delta_{\mathrm{m}}^{2} \kappa_{\mathrm{a}}-\mathrm{g}_{1}^{2} \kappa_{\mathrm{m}}+\mathrm{g}_{2}^{2} \kappa_{\mathrm{m}}+\Delta_{\mathrm{a}}^{2} \kappa_{\mathrm{m}}\nonumber+\frac{1}{4} \kappa_{\mathrm{a}}^{2} \kappa_{\mathrm{m}}+\frac{1}{4} \kappa_{\mathrm{a}} \kappa_{\mathrm{m}}^{2},\nonumber\\{H}_{0}&=&\mathrm{g}_{1}^{4}-2 \mathrm{~g}_{1}^{2} \mathrm{~g}_{2}^{2}+\mathrm{g}_{2}^{4}-2 \mathrm{~g}_{1}^{2} \Delta_{\mathrm{a}} \Delta_{\mathrm{m}}-2 \mathrm{~g}_{2}^{2} \Delta_{\mathrm{a}} \Delta_{\mathrm{m}}+\Delta_{\mathrm{a}}^{2} \Delta_{\mathrm{m}}^{2}+\frac{1}{4} \Delta_{\mathrm{m}}^{2} \kappa_{\mathrm{a}}^{2}-\frac{1}{2} \mathrm{~g}_{1}^{2} \kappa_{\mathrm{a}} \kappa_{\mathrm{m}}+\frac{1}{2} \mathrm{~g}_{2}^{2} \kappa_{\mathrm{a}} \kappa_{\mathrm{m}}+\frac{1}{4} \Delta_{\mathrm{a}}^{2} \kappa_{\mathrm{m}}^{2}+\frac{1}{16} \kappa_{\mathrm{a}}^{2} \kappa_{\mathrm{m}}^{2}.
\end{eqnarray}

 Based on the Fourier transform for the operator ${\hat{O}(\omega)=\frac{1}{2\pi}\int_{-\infty}^{\infty}dtO(t)e^{i\omega t}}$,   the fluctuation operator in the time domain can be transferred to the frequency domain, hence from the input-output relationship $\delta\hat{P}^{out}_a=\sqrt{\kappa_a}\delta \hat{P}_a-\hat{p}^{in}_a$, one can obtain the quadrature component of the phase as
	\begin{align}\label{14}
		\delta\hat{P}^{\rm{out}}_a(\omega)=&M_1(\omega)\hat{x}^{\prime \rm{in}}_m(\omega)+M_2(\omega)\hat{p}^{\prime \rm{in}}_m(\omega)+M_3(\omega)\hat{x}^{\rm{in}}_a(\omega)+M_4(\omega)\hat{p}^{\rm{in}}_a(\omega),
	\end{align}
where $\hat{x}^{\prime\rm{in}}_m(\omega)=\hat{x}^{\rm{in}}_m (\omega)+\epsilon\sqrt{\frac{1}{2\kappa_m}}{i} [B_{ex}(\omega+\omega_d)-B_{ex}(\omega-\omega_d)]$ and $\hat{p}^{\prime\mathrm{in}}_m(\omega)=\hat{p}^{\rm{in}}_m (\omega)+\epsilon\sqrt{\frac{1}{2\kappa_m}}[B_{ex}(\omega+\omega_d)+B_{ex}(\omega-\omega_d)]$ are the magnon orthogonal fluctuation operator including the external detected magnetic field, and
		\begin{align}\label{15}
	    {M_1}(\omega)=&\frac{\chi_{a} \chi_{m}^{\prime} \sqrt{\kappa_{a} \kappa_{m}}\left\{\left(g_{1}+g_{2}\right)+\chi_{a} \chi_{m}\left(g_{1}-g_{2}\right)[\Delta_{a} \Delta_{m}-\left(g_{1}+g_{2}\right)^{2}]\right\}}{1+2 \chi_{a} \chi_{m}^{\prime}\left(g_{2}^{2}-g_{1}^{2}\right)+\chi_{a}^{2}\left\{\Delta_{a}^{2}+\chi_{m}\chi^{\prime}_{m}
	    \left[(g_{1}^{2}-g_{2}^{2})^{2}-2(g_{1}^{2}+g_{2}^{2})\right]\right\}},\nonumber\\
		{M_2}(\omega)=&\frac{\chi_{a} \chi_{m}^{\prime} \sqrt{\kappa_{a} \kappa_{m}}\left[\chi_m(g_1+g_2)\Delta_{m}-\chi_a(g_1-g_2)\Delta_{a}\right]}{1+2 \chi_{a} \chi_{m}^{\prime}\left(g_{2}^{2}-g_{1}^{2}\right)+\chi_{a}^{2}\left\{\Delta_{a}^{2}+\chi_{m}\chi^{\prime}_{m}
			\left[(g_{1}^{2}-g_{2}^{2})^{2}-2(g_{1}^{2}+g_{2}^{2})\right]\right\}},\nonumber\\
	    {M_3}(\omega)=&\frac{\chi_{a}^2[\Delta_{a}-\chi_{m}\chi_{m}^{\prime}(g_{1}+g_{2})^2\Delta_{m}]}{1+2 \chi_{a} \chi_{m}^{\prime}\left(g_{2}^{2}-g_{1}^{2}\right)+\chi_{a}^{2}\left\{\Delta_{a}^{2}+\chi_{m}\chi^{\prime}_{m}
	    	\left[(g_{1}^{2}-g_{2}^{2})^{2}-2(g_{1}^{2}+g_{2}^{2})\right]\right\}},\nonumber\\
		{M_4}(\omega)=&\frac{1-\chi_{a}\left[2 \chi_{m}^{\prime}(g_{1}^{2}-g_{2}^{2})+\kappa_{a}\right]+x_{a}^{2}\left\{\Delta_{a}^{2}+\chi_{m}^{\prime} \chi_{m}(g_{1}^{2}-g_{2}^{2})-\chi_{m}^{\prime}\left[2 \chi_{m}(g_{1}^{2}+g_{2}^{2})\Delta_{a}\Delta_{m}+(g_{2}^{2}-g_{1}^{2}) \kappa_{a}\right]\right\}}{1+2 \chi_{a} \chi_{m}^{\prime}\left(g_{2}^{2}-g_{1}^{2}\right)+\chi_{a}^{2}\left\{\Delta_{a}^{2}+\chi_{m}\chi^{\prime}_{m}\left[(g_{1}^{2}-g_{2}^{2})^{2}-2(g_{1}^{2}+g_{2}^{2})\right]\right\}},
	\end{align}
with $\chi_a(\omega)=\frac{1}{\kappa_a/2-i\omega}$ and $\chi_m(\omega)=\frac{1}{\kappa_m/2-i\omega}$ being the susceptibility of the cavity field and the magnon mode, and $\chi_{m}^{\prime}=\frac{1}{1/\chi_m+\Delta_m^2}$ denoting the effective susceptibility of the magnon mode.
	\end{widetext}
To achieve weak magnetic field sensing, we need to use a homodyne detection device to detect the phase output symmetrical spectrum density, defined as \cite{PhysRevA.100.023815} 
\begin{align}\label{16}
	Y_{{\mathrm{out}}}(\omega)=&\frac{1}{2}\int d\omega^{\prime}e^{i(\omega+\omega^{\prime})t}\langle \delta P^{\rm{out}}_a(\omega)\delta P^{\rm{out}}_a(\omega^{\prime})\nonumber\\
	&+\delta P^{\rm{out}}_a(\omega^{\prime})\delta P^{\rm{out}}_a(\omega)\rangle.
\end{align}
According to Eq. (\ref{7}), the cavity field phase output spectrum density of the system can be written as
\begin{align}\label{17}
	Y_{{\rm{out}}}(\omega)&=(\bar{n}_a+\frac{1}{2})[\vert {M_4}(\omega)\vert ^2]\notag\\&+\vert {M_1}(\omega)\vert ^2[(\bar{n}_m+\frac{1}{2})+S_{B_{ex}(\omega)}(\omega)] ,
\end{align}
where $S_{B_{ex}}$ is the signal spectral density of the external magnetic field corresponding to the magnon mode amplitude orthogonal component, respectively.
From the expression of the output spectrum, it can be seen that the first term represents the contribution of the microwave cavity field, while the second term represent the contribution of the YIG probe and the detected magnetic field signal.
Next, in order to better measure the performance of weak magnetic field sensing, we define	
\begin{align}
	R_B(\omega)&={\partial Y_{\rm{out}}(\omega)/\partial S_{B_{ex}(\omega)}(\omega)}=\vert {M_1}(\omega)\vert ^2,\label{18} \\
	N_{\rm{\rm{ad}}}(\omega)&=(\bar{n}_a+\frac{1}{2})\frac{ [\vert {M_4}(\omega)\vert ^2]}{{\partial Y_{\rm{out}}(\omega)/\partial S_{B_{ex}(\omega)}(\omega)}}\notag\\&=(\bar{n}_a+\frac{1}{2})\frac{ [\vert {M_4}(\omega)\vert ^2]}{\vert {M_1}(\omega)\vert ^2},\label{19}
\end{align}
represent the response to external magnetic signals and the additional noise including cavity field thermal noise and quantum noise in weak magnetic field sensing, respectively.
From Eq. (\ref{17}), one can see that $R_B(\omega)>1$ indicates the signal amplification, and the smaller the additional noise $N_{\rm{\rm{ad}}}(\omega)$ of the system is, the easier it is to detect the signal of the external magnetic field. Besides, when $N_{\rm{\rm{ad}}}(\omega)<1/2$, we say that additional noise is below the SQL \cite{PhysRevA.100.023815}. Therefore, reducing the additional noise below the SQL will improve the sensitivity of weak magnetic sensing. In this sense, we won't consider the specific expression of the detected magnetic field signal spectrum but focus on reducing the additional noise and enhancing the response.	
	\section{\label{sec4} WEAK MAGNETIC FIELD SENSING}

	\begin{figure}
		\includegraphics[width=10.4cm,height=8.0cm]{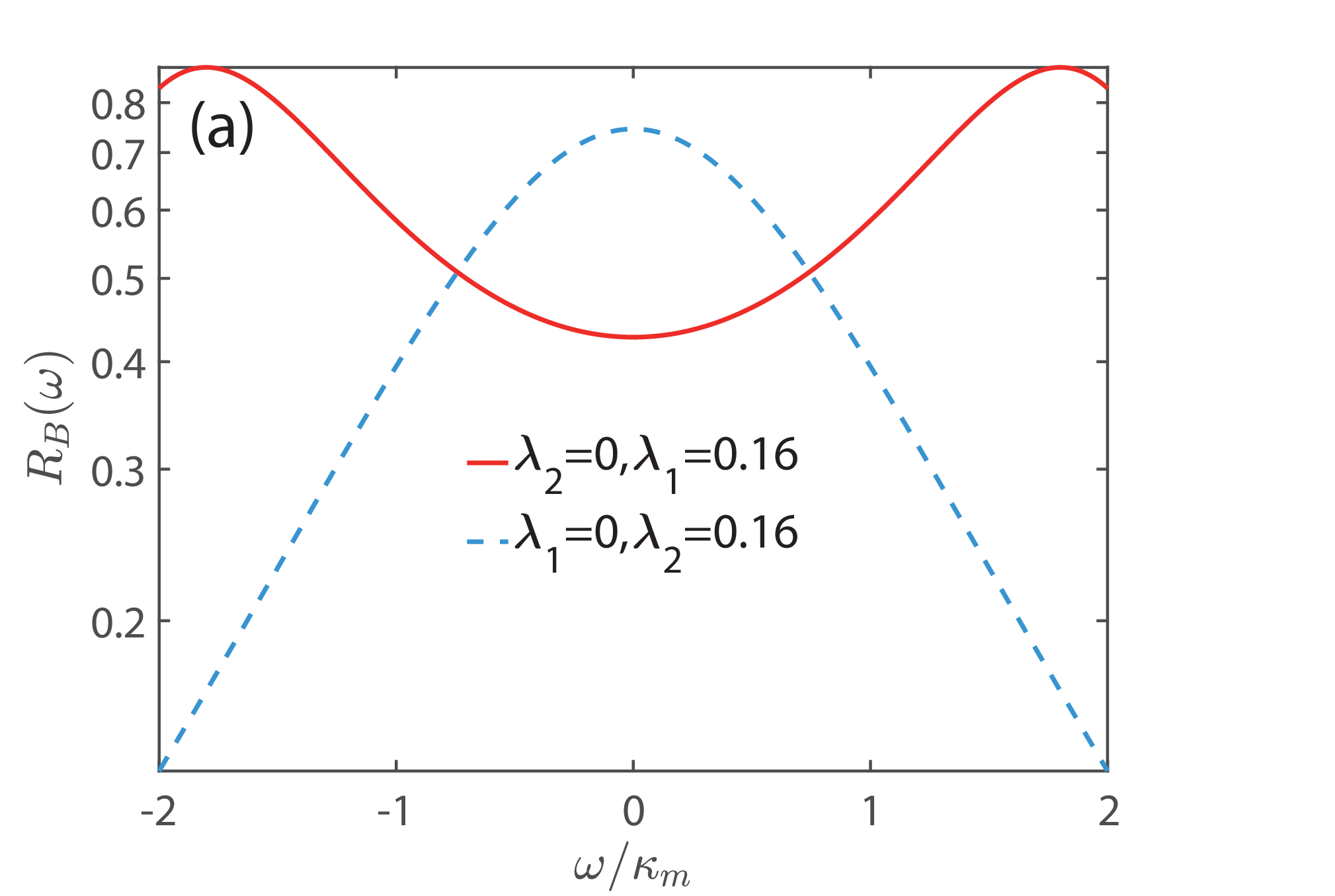}
		\hspace{10mm}
		\includegraphics[width=10.4cm,height=8.0cm]{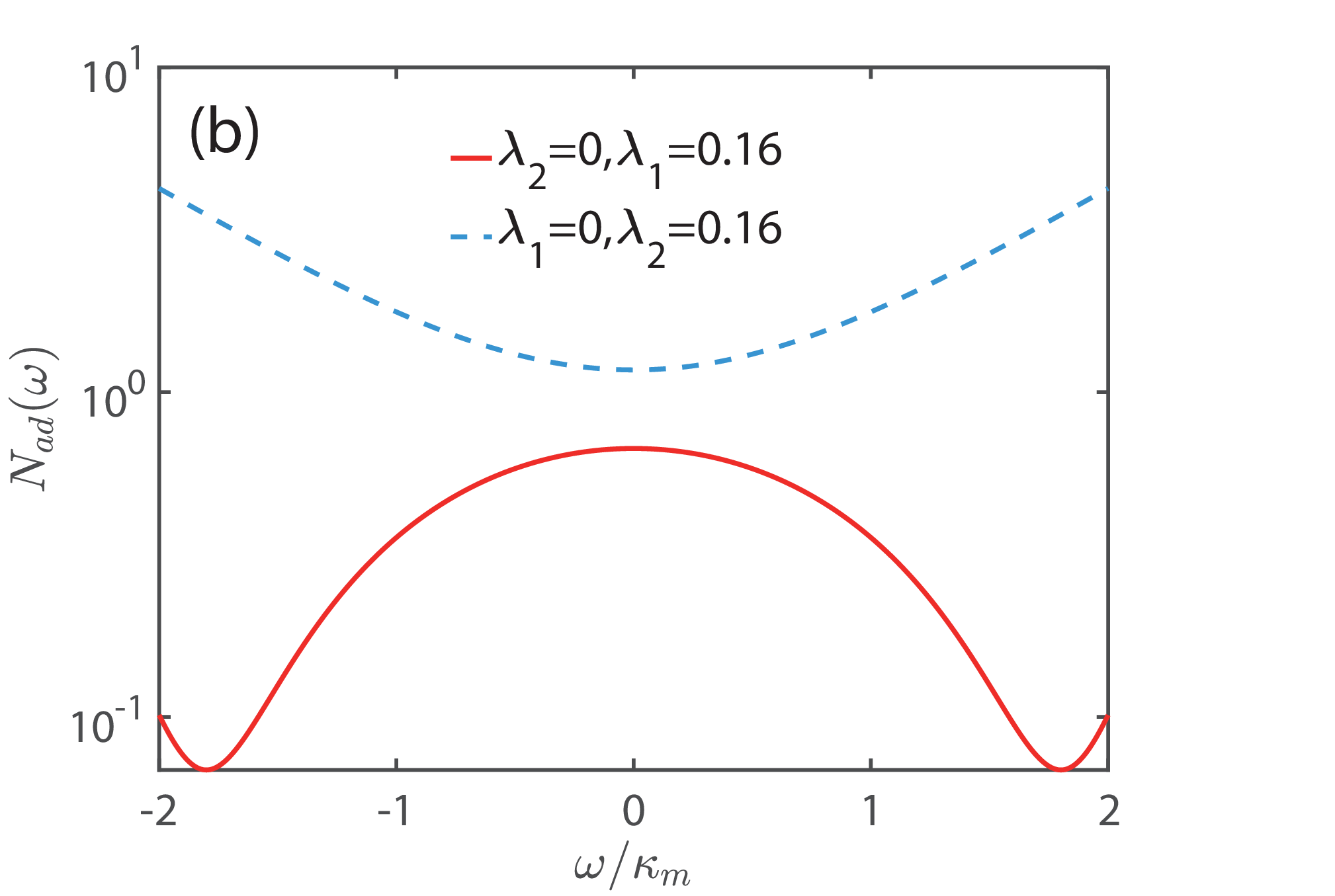}
		\caption{(a) The response $R_B$ of the system as a function of the normalized frequency $\omega/\kappa_{m}$. (b) The additional noise $N_{ad}$ as a function of normalized frequency $\omega/\kappa_{m}$. The red solid line represents the case of only rotating wave term interaction, while the blue dashed line represents the case of only anti-rotating wave term interaction. The experimental parameters related to Fig. 2(a) and Fig. 2(b) are: $\omega_{m}/2\pi=37.5~\rm {GHz}$, $g=10 ^ {-2}\omega_{ m }$, $\Delta_a=\Delta_{m}=0$, $\kappa_{m}/2\pi=15~\rm {MHz}$, $\kappa_{a}/2\pi=33~\rm {MHz}$. In addition, we set the environment temperature to  $\rm 50~\rm {mK}$, which is unaffected by thermal noise. }
		\label{Fig2}
	\end{figure}
	\begin{figure}
		\includegraphics[width=9.3cm,height=8.0cm]{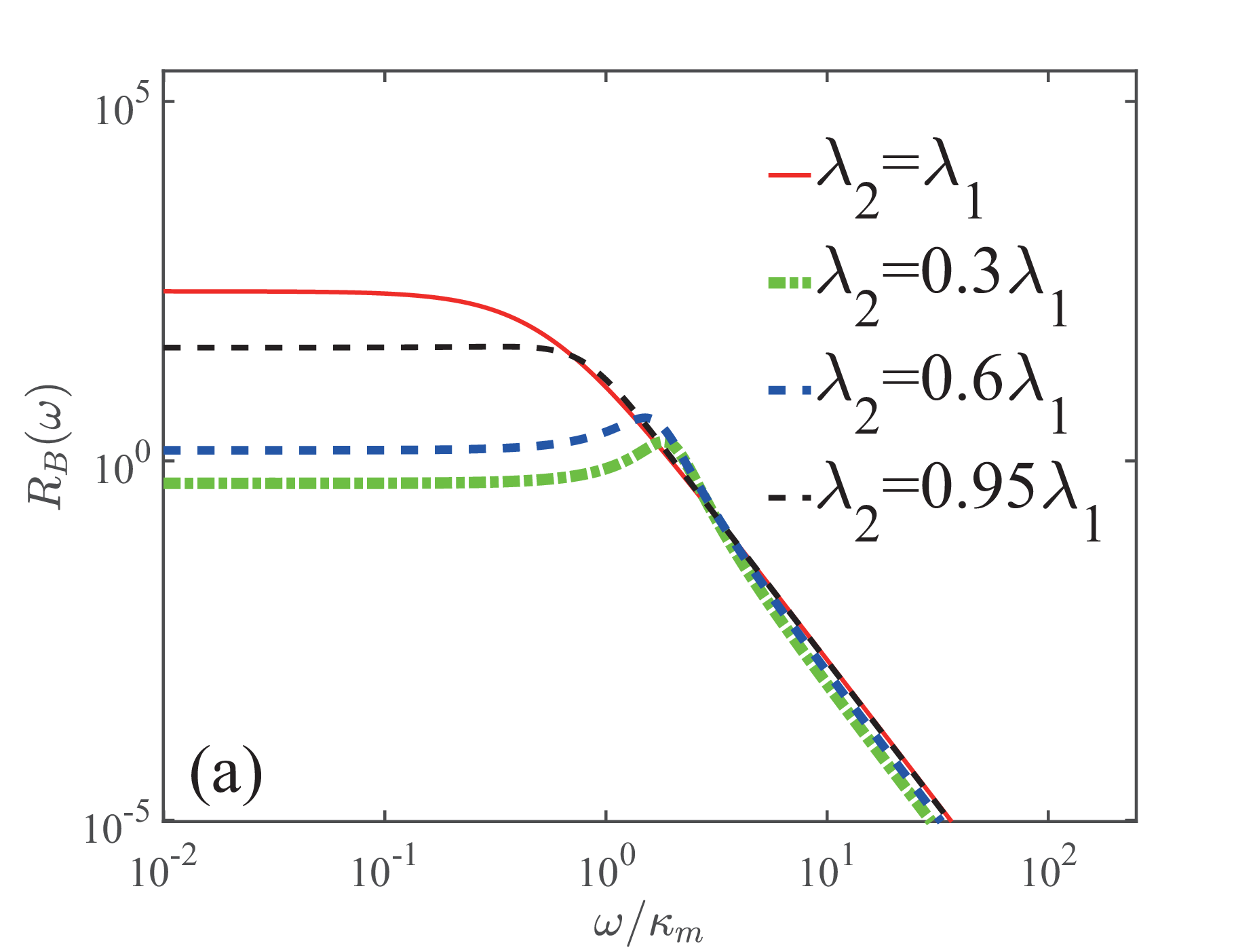}
		\hspace{10mm}
	\includegraphics[width=9.3cm,height=8.0cm]{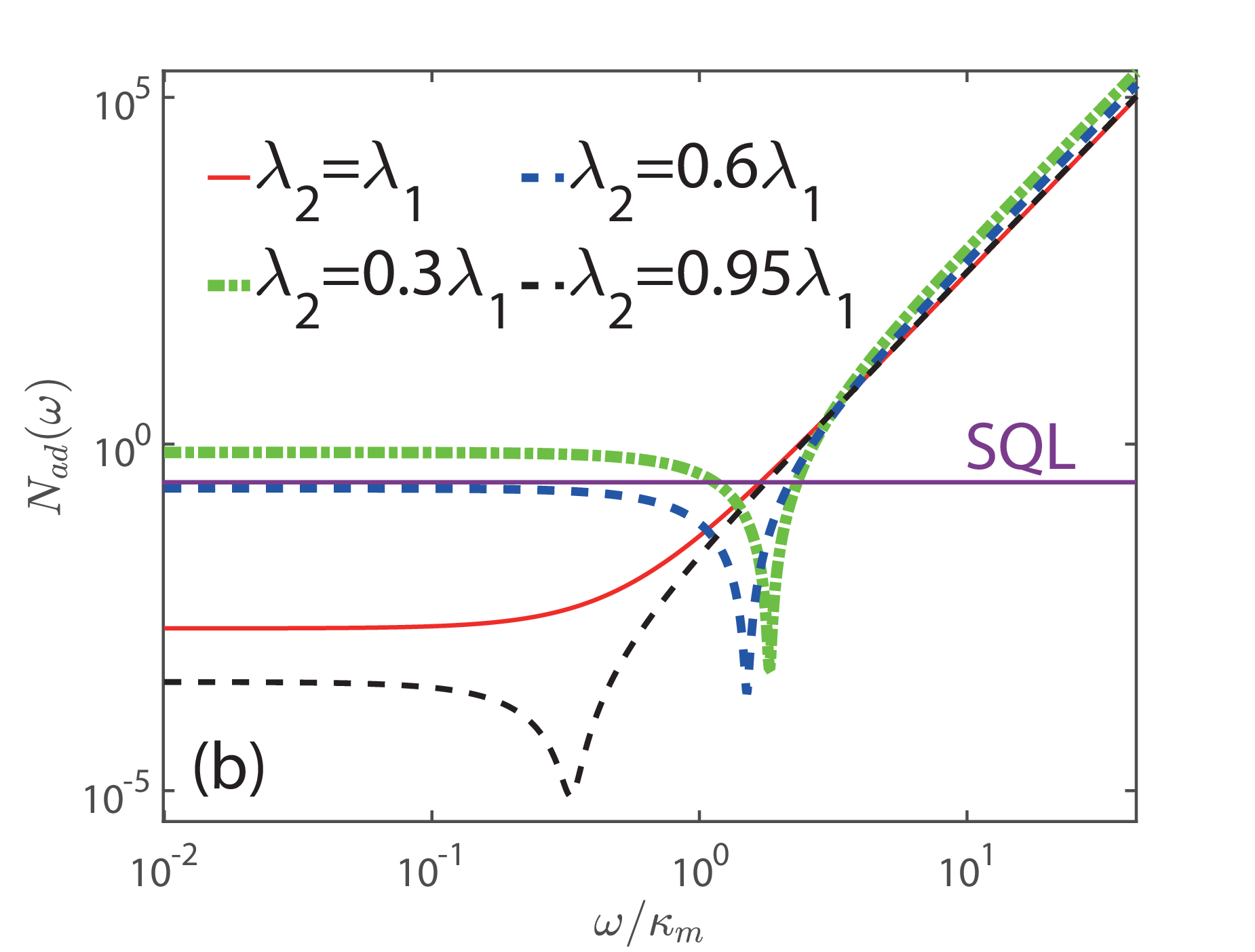}
		\caption{(a) The response $R_B$ of the system as a function of the normalized frequency $\omega/\kappa_{m}$, with logarithmic scale on the x-axis. (b) The additional noise $N_{ad}$ as a function of normalized frequency $\omega/\kappa_{m}$,with logarithmic scale on the x-axis. $\lambda_2$ is set to 0, 0.3$\lambda_1$, 0.6$\lambda_1$, and 0.95$\lambda_1$. Other parameters are: $\omega_{m}/2\pi=37.5~\rm {GHz}$, $g=10 ^ {-2}\omega_{ m }$, $\lambda_1=0.16$, $\Delta_a=\Delta_{m}=0$, $\kappa_{m}/2\pi=15~\rm {MHz}$, $\kappa_{a}/2\pi=33~\rm {MHz}$. In addition, we set the environment temperature to $\rm 50~\rm {mK}$, which is almost  unaffected by thermal noise.}
		\label{Fig3}
	\end{figure}
	\begin{figure}
\includegraphics[width=10.5cm,height=8.0cm]{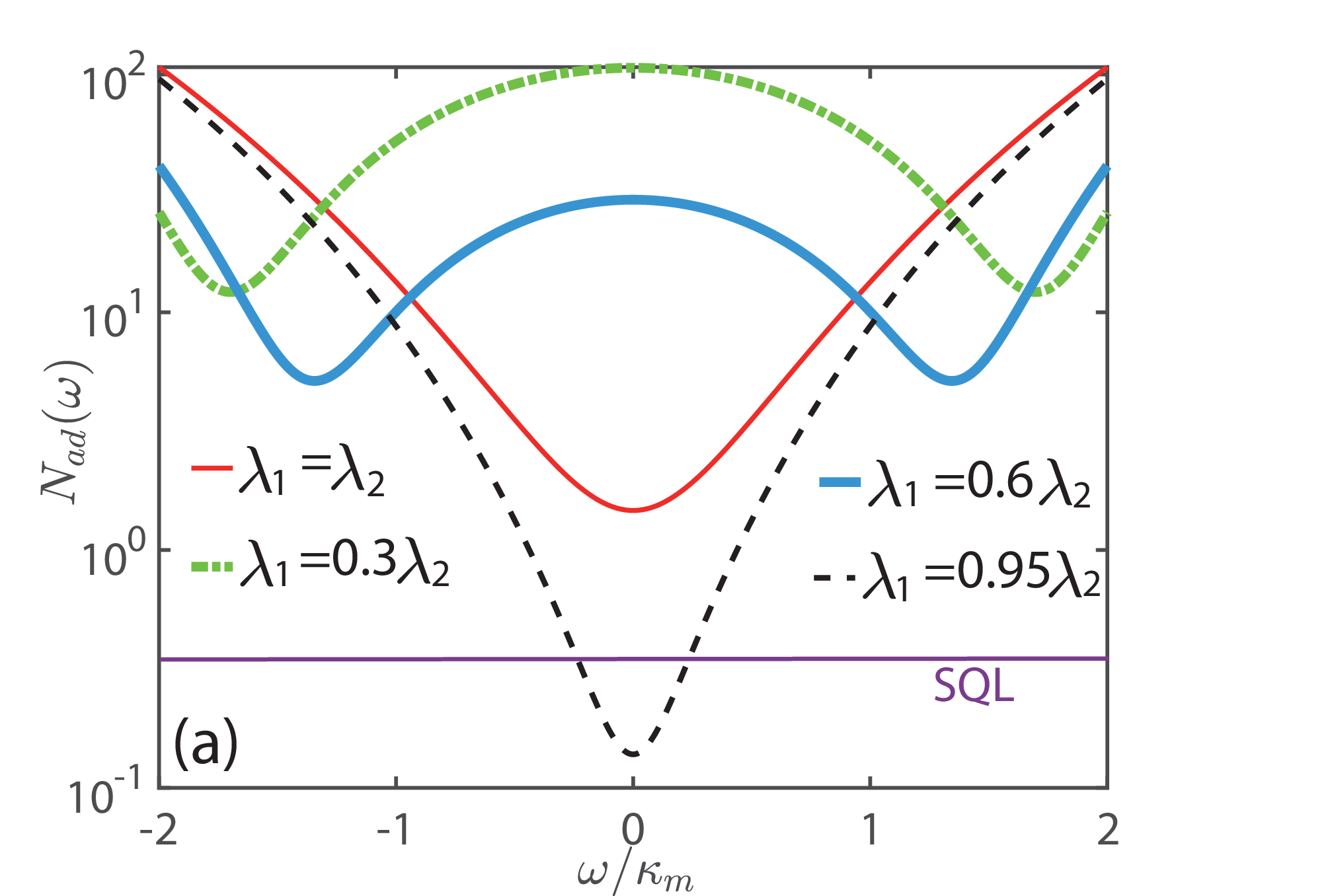}
	\hspace{10mm}
\includegraphics[width=9.6cm,height=8.4cm]{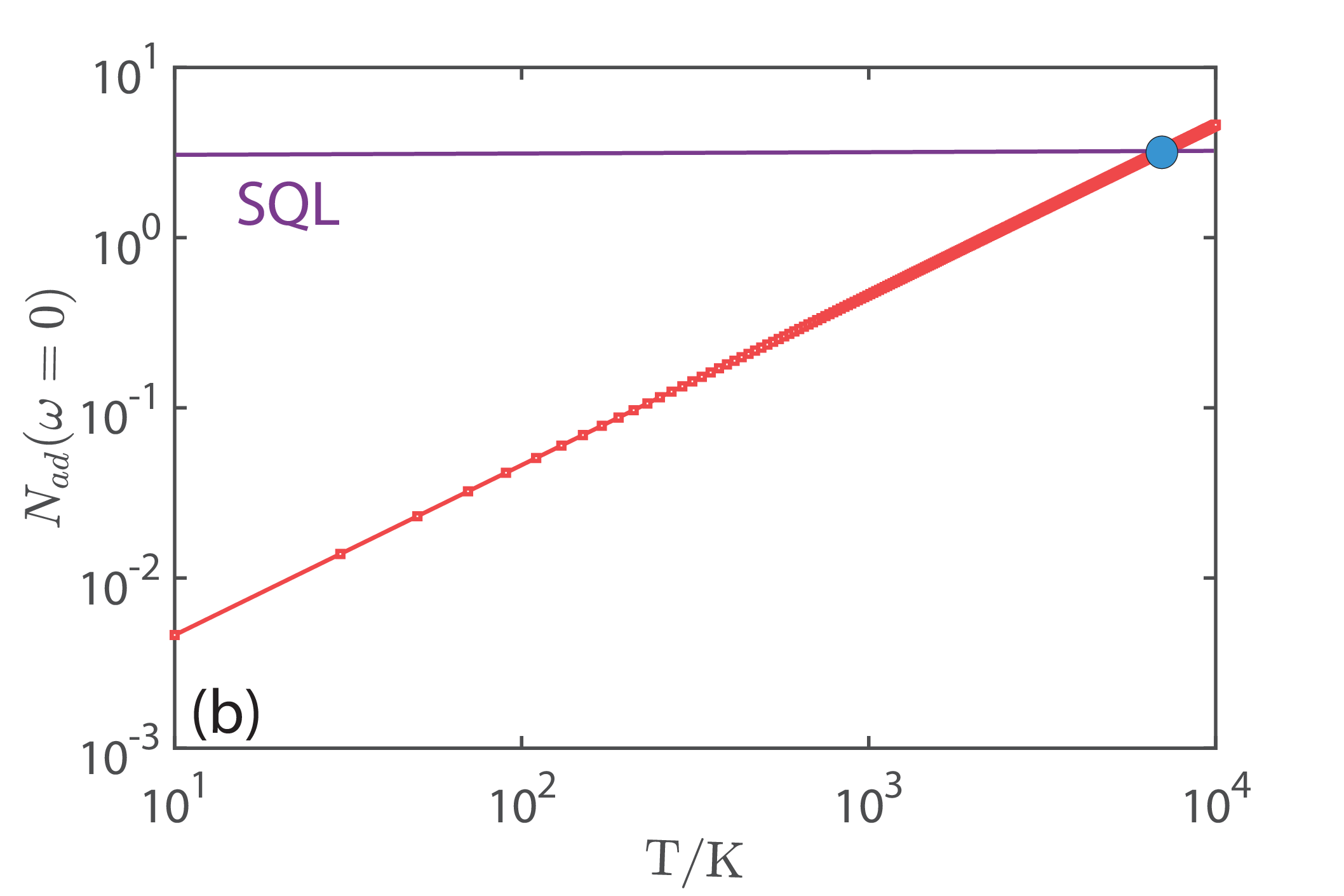}
	\caption{(a) The additional noise $N_{ad}$ as a function of normalized frequency $\omega/\kappa_{m}$. Unlike Fig. \ref{Fig3}(b), we set the temperature to room temperature, i.e. $T=300~\rm {K}$. (b) The additional noise $N_{ad} (\omega=0)$ as a function of temperature $T$. the purple solid line represents the  SQL. The other experimental parameters related to Fig. 4(a) and Fig. 4(b) are: $\omega_{m}/2\pi=37.5~\rm {GHz}$, $g=10 ^ {-2}\omega_{ m }$, $\lambda_1=0.16$, $\Delta_a=\Delta_{m}=0$, $\kappa_{m}/2\pi=15~\rm {MHz}$, $\kappa_{a}/2\pi=33~\rm {MHz}$.}
	\label{Fig4}
\end{figure}	
In this section, we provide numerical results of the performance of our weak magnetic sensing scheme. Our scheme can greatly relax the requirements for cavity magnetic coupling strength and achieve additional noise suppression of the same order of magnitude without dual frequency modulation under the super strong coupling mechanism or deep strong coupling mechanism. The scheme without dual frequency modulation is recently proposed by M S. Ebrahimi $et$ $al$ \cite{PhysRevA.103.062605}. Therefore, we reproduced some of their results and compared them with our scheme. In addition, we have set the following feasible experimental parameters \cite{PhysRevLett.113.156401}: $\omega_{m}/2\pi=37.5~\rm {GHz}$, $g=10 ^ {-2}\omega_{ m }$, $\Delta_a=\Delta_{m}=0$, $\kappa_{m}/2\pi=15~\rm {MHz}$, $\kappa_{a}/2\pi=33~\rm {MHz}$.

First, we'd like to study how the rotating-wave interaction and anti-rotating wave interaction affect weak magnetic field sensing. We set $\Delta_a=\Delta_m=0$, which can be achieved by adjusting the bias field and the frequency of the external driving microwave field. This setting can decouple the two orthogonal components of the magnon mode and the orthogonal components of the cavity field, thereby improving the sensitivity of the measurement. We plot the additional noise of weak magnetic field sensing and the response to external magnetic signals versus the frequency in Fig. \ref{Fig2}. The red solid line corresponds to the presence of the only rotating wave term, and the blue dashed line indicates the only anti-rotating wave term present. Fig. \ref{Fig2}(a) shows that neither type of interaction alone cannot increase the system's response to external magnetic field signals, since the curves are below $10^0$. Fig. \ref{Fig2}(b) shows that under resonance conditions ($\omega\approx0$), the additional noise suppression level of the rotating wave term is greater than that of the non rotating wave term, and it also has great noise suppression performance in the non-resonant region. This result indicates that the interaction of rotating wave terms is beneficial for improving the sensitivity of detection, which can be understood as the result of a beam splitter type interaction between the magnon and the cavity photon. That is, when the external magnetic field affects the magnon, coherent energy exchange occurs between the magnon and the photon, which is more conducive to optical field readout and thus improves sensitivity.

In Fig. \ref{Fig3}(a), we plot the response of the system with additional noise under different $\lambda_2/\lambda_1$ with both the rotating and non-rotating wave terms present. The red curve corresponds to  $\lambda_2=\lambda_1$, namely, the rotating wave term and the non-rotating wave term has the same weight in the Hamiltonian.  One can find that with the relative weight ($\lambda_2/\lambda_1$) of the non-rotating wave term increasing, the response of the system is correspondingly improved in the resonance region. This result indicates that a single parametric amplification interaction (non-rotating wave term) cannot amplify the signal to be detected, and it must coexist with the beam splitter type interaction (rotating wave term). Fig. \ref{Fig3}(b) shows the additional noise of the system with different $\lambda_2/\lambda_1$. At the low environment temperature of $50~\rm {mK}$ implying the low environment thermal noise, it can be found that under resonance condition ($\omega\approx0$), the best suppression of additional noise is not under the condition of $\lambda_2=\lambda_1$, but at $\lambda_2=0.95\lambda_1$. At $\omega\approx0.33\kappa_m$, there is a valley region,  where the additional noise suppression effect is one order of magnitude stronger than the case of $\lambda_2=\lambda_1$. In addition, for $\lambda_2/\lambda_1=0.3,0.6$, the suppression effect isn't apparent even though both have a suppression valley.
	
To emphasize the performance of this sensor in the presence of environmental thermal noise, in Fig. \ref{Fig4}(a), we study the additional noise of the system versus the normalized frequency at room temperature ($T=300~\rm K$). One can find that at $\lambda_2=0.95\lambda_1$, the additional noise can still be suppressed below the SQL ($N_{ad}(\omega)<\frac{1}{2}$) in the  $\omega \approx 0$.  Especially in  Fig. \ref{Fig4}(b), we plot the curves of the additional noise with temperature, which indicates that in a wide temperature range, the additional noise can be suppressed below the SQL. In particular, our scheme can extremely strongly suppress the additional noise under the strong coupling regime instead of the ultra-strong or deep strong coupling required for the anti-rotating wave term interaction.  
\begin{figure}
    \includegraphics[width=10.3cm,height=8.0cm]{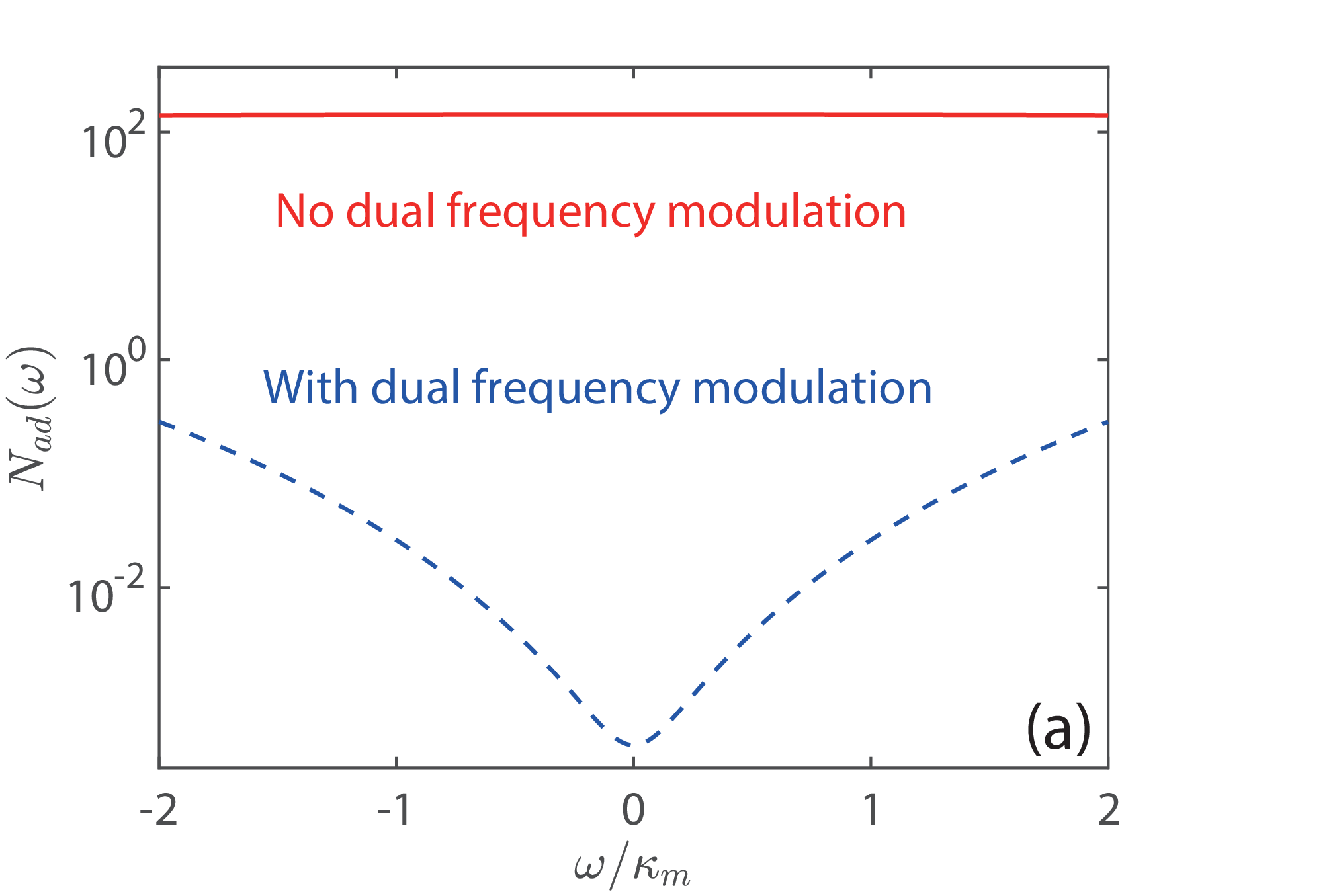}
	\hspace{10mm}
	\includegraphics[width=10.3cm,height=8.0cm]{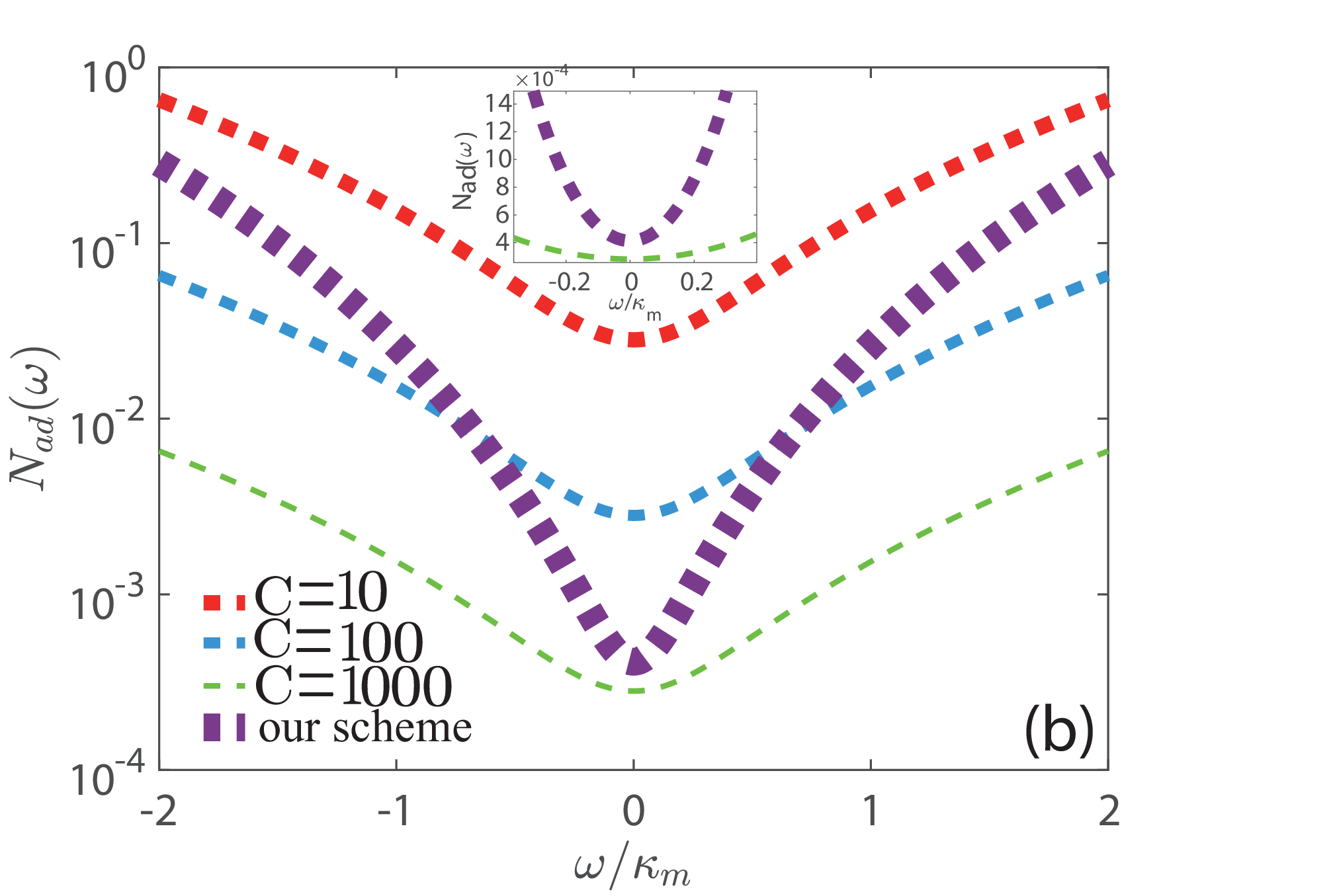}
	\caption{(a) The additional noise $N_{ad}$ of the system is taken as a function of the normalized frequency $\omega/\kappa_{m}$. The blue dashed line represents the presence of a dual frequency modulated magnetic field, the red solid line represents the absence of a dual frequency modulated magnetic field. (b) The additional noise $N_{ad}$ as a function of normalized frequency $\omega/\kappa_{m}$ under different electromagnetic cooperativity $C$. Specifically, the red, green, and blue dashes represent the corresponding cases of $C=10$, $C=100$, and $C=1000$ in the ultra-strong coupling mechanism scheme. The purple dashed line represents our dual frequency modulation bias field scheme. The other experimental parameters related to Fig. 5(a) and Fig. 5(b) are: $\omega_{m}/2\pi=37.5~\rm {GHz}$, $g=10 ^ {-2}\omega_{ m }$, $\lambda_1=0.16$, $\lambda_2=0.95\lambda_1$, $\Delta_a=\Delta_{m}=0$, $\kappa_{m}/2\pi=15~\rm {MHz}$, $\kappa_{a}/2\pi=33~\rm {MHz}$, $\rm T=50~\rm {mK}$.}
	\label{Fig7}
\end{figure}

Next, we compared the additional noise suppression performance of our scheme with the scheme without dual frequency bias field modulation. In In Fig. \ref{Fig7}(a), the red solid line represents the scheme without dual frequency modulation, and the blue dashed line represents the additional noise suppression result of our scheme. It can be seen that the red solid line does not have a good additional noise suppression effect, which is the same as the result of Ref\cite{PhysRevA.103.062605}. However, our scheme can achieve the additional noise suppression effect in the resonance region that exceeds the scheme without dual frequency modulation by nearly five orders of magnitude.

Before the end, we'd like to emphasize that dual frequency bias field driving is a key to our scheme. Next, we will pay attention to the case of the {\it{ultra-strong or deep-strong}} coupling scheme. The zero detuning conditions ($\Delta_m=\Delta_a=0$) for the ultra-strong coupling scheme are given as   \cite{PhysRevA.103.062605}  
	\begin{eqnarray}
		R_{\rm B2}(\omega) &=& 4\kappa_a^2\kappa_m^2 C \big \vert \frac{\chi_{m1}(\omega)}{2i\omega+\kappa_a} \big \vert ^2 ,\label{31} \\
		N_{\rm{ad2}}(\omega) & = & \frac{\big \vert 1+\frac{2\kappa_a}{2i\omega+\kappa_a}\big \vert^2(\bar{n}_a+\frac{1}{2})}
		{4\kappa_a^2\kappa_m^2 C \big \vert \frac{\chi_{m1}(\omega)}{2i\omega+\kappa_a} \big \vert ^2},\label{32}
	\end{eqnarray}
where $\chi_{m1}(\omega)= \frac{1}{\kappa_m/2+i\omega}$ is the susceptibility, $C$ is the electromagnetic cooperativity, with a range of 1-1000.
In Fig. \ref{Fig7}(b), we compare our scheme with the scheme under the super-strong coupling mechanism. It can be seen that our our ability to suppress additional noise can even reach the level of the ultra-strong coupling scheme with extremely high electromagnetic cooperativity. Specifically, in Fig. \ref{Fig7}(b). One can find that the noise suppression effect of our scheme near $\omega\approx0$ is significantly stronger than that of $C=10,100$. Moreover, our scheme can achieve noise suppression of the same order of magnitude under the condition of $C=1000$, as shown in the small figure in Fig. \ref{Fig7}(b). When $\omega\approx0$, the additional noise corresponding to the green dashed line is 0.0002, and our scheme can achieve 0.0004. This demonstrates the potential of our scheme to achieve similar noise suppression capabilities as the ultra-strong coupling scheme under a strong coupling mechanism without any parameter adjustments. The comparison in Fig. \ref{Fig7}(b) indicates that our scheme can achieve high sensitivity without the necessity of experimental conditions associated with an ultra-strong coupling mechanism. Of course, suppose our scheme can be implemented for magnetic field sensing under the ultra-strong coupling mechanism. In that case, it is natural that the sensitivity and the ability of signal amplification will be further enhanced. This sufficiently demonstrates the superiority of our scheme.  

\section{\label{sec5} DISCUSSION and CONCLUSIONS}
	
In summary, we proposed a scheme for weak magnetic field sensing using dual frequency bias field modulation. The introduction of the dual bias field modulation can adjust the proportion of the rotating and the anti-rotating wave type interaction in the weak magnetic field sensing system, which help to reveal the crucial roles of the two types of interactions played in the weak magnetic field sensing. It is found that the anti-rotating wave term can amplify the magnon mode signal, but this amplification effect must coexist with the rotating wave term. Besides, The distinct advantage is that our scheme can achieve a more sensitive and temperature-robust weak magnetic field sensing with additional noise , but it does not require an ultra- or a deep-strong coupling mechanism. In this sense, it reduces the difficulty of experimental implementation. In addition,, We compared our scheme with previous schemes that required ultra- strong coupling or deep-strong coupling mechanisms and found that in terms of additional noise suppression, our scheme can achieve the additional noise suppression of the same order of magnitude as the previous scheme. This indicates that we can not only relax the experimental conditions, but also have no decrease in sensitivity. Therefore, it can be said that our scheme achieves highly sensitive weak magnetic field sensing. Moreover, although we have achieved the temperature robustness of the additional noise, the thermal noise of the magnon mode is a significant problem, and we will study how to reduce the thermal noise of the magnon in the future.

\acknowledgments
	This work was supported by the National Natural Science Foundation of China
	under Grant No.12175029, No. 12011530014, and No.11775040, and the Key
	Research and Development Project of Liaoning Province, under grant
	2020JH2/10500003.
\bibliography{magnetometry}

\end{document}